\documentclass[11pt,twoside]{article}
\usepackage{asp2010}

\resetcounters

\begin{document}

\title{Reflex: Scientific Workflows for the ESO Pipelines}
\author{Pascal~Ballester$^1$, Daniel~Bramich$^1$, Vincenzo~Forchi$^1$, Wolfram~Freudling$^1$, Cesar~Enrique~Garcia-Dab{\'o}$^2$, Maurice~klein~Gebbinck$^1$, Andrea~Modigliani$^1$ Martino~Romaniello$^1$}
\affil{$^1$European Southern Observatory – Karl-Schwarzschild-Str. 2 D-85748 Garching bei München, Germany}
\affil{$^2$FRACTAL SLNE,  Tulip{\'a}n 2, portal 13, 1. A E-28231 Las Rozas de Madrid, Spain}

\begin{abstract}
The recently released Reflex scientific workflow environment supports the interactive execution 
of ESO VLT data reduction pipelines. Reflex is based upon the Kepler workflow engine, and provides 
components for organising the data, executing pipeline recipes based on the ESO Common Pipeline Library, 
invoking Python scripts, and constructing interaction loops. Reflex will greatly enhance the quick 
validation and reduction of the scientific data. In this paper we summarize the main features of Reflex, 
and demonstrate as an example its application to the reduction of echelle UVES data. 
\end{abstract}

\section{Reflex, a data processing workflow environment based on Kepler}

The ESO Recipe Flexible Execution Workbench (Reflex) is an environment which allows an easy and flexible way to execute VLT pipelines. It is built using the Kepler workflow engine (https://kepler-project.org). Reflex allows the users to process their scientific data in the following steps: 

\begin{itemize}
\item Associate scientific files with required calibrations, 
\item Choose datasets to be processed, 
\item Execute several pipeline recipes in a logical sequence,
\item Inspect intermediate products, and 
\item Interactively change recipe parameters.
\end{itemize}

A workflow accepts science and calibration data, as delivered to Principal Investigators (PIs) in the form of PI-Packs or as
downloaded from the archive, and organises them into groups of files called LoSOs (List of Science Observations),
where each LoSO contains one science object observation and all associated raw calibrations
required for a successful data reduction. The data organisation process is fully automatic, which is a major
time-saving feature provided by the software. The LoSOs selected by the user for reduction are fed through
the workflow which executes the relevant pipeline recipes (or stages) in the correct order, providing optional
user interactivity at key data reduction points with the aim of enabling the iteration of certain recipes in order
to obtain better results. Full control of the various recipe parameters is available within the workflow,
and the workflow deals automatically with optional recipe inputs via built-in conditional branches. Additionally,
the workflow stores the reduced final data products in a logically organised directory structure and
employing user-configurable file names.

\begin{center}
\begin{figure}[!ht]
\includegraphics[width=130mm,height=90mm]{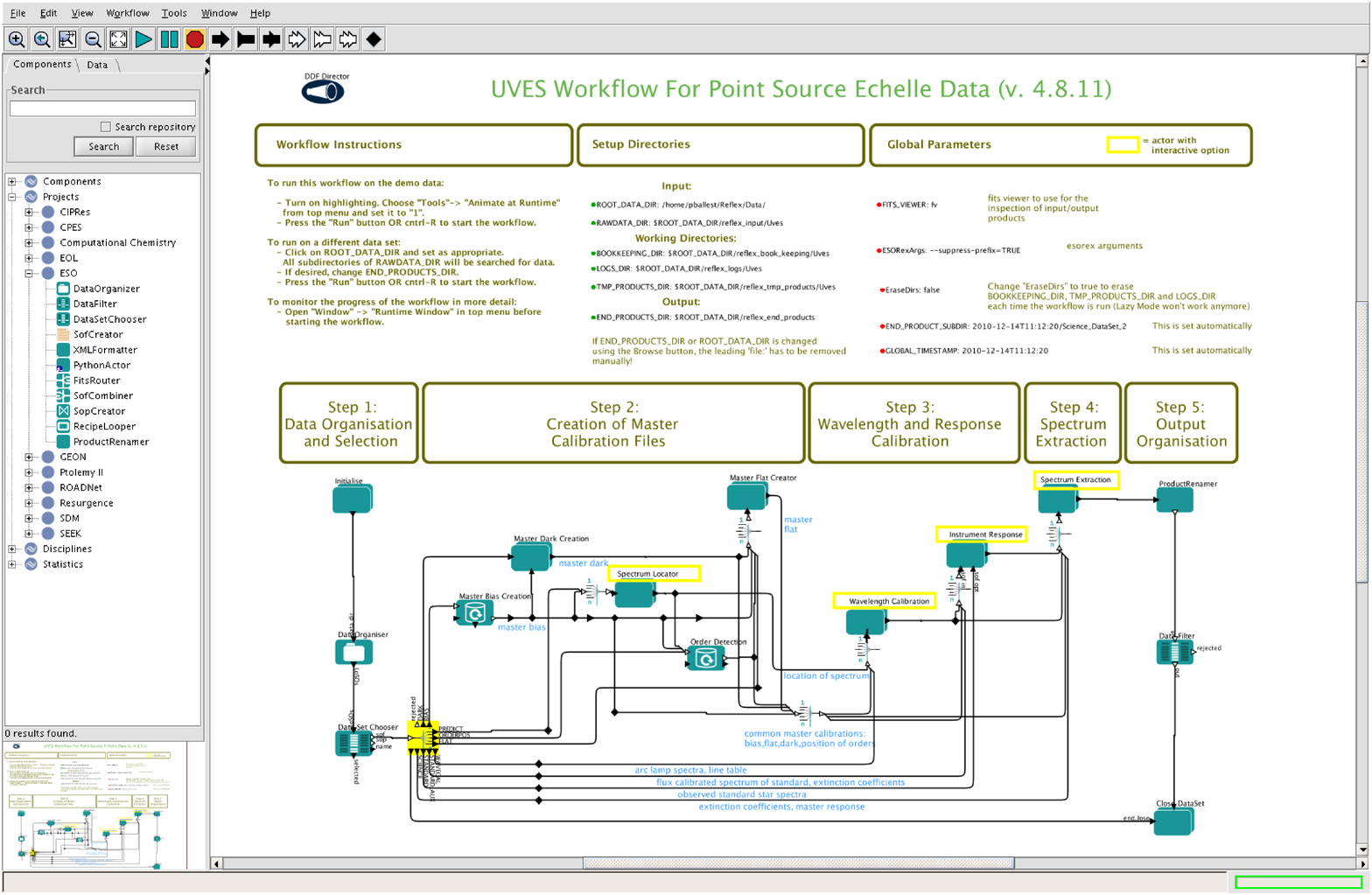}
\caption{The main workflow window for the UVES pipeline. Annotations and top-level parameters of the workflow are shown to the user, while the logic pertaining to individual reduction steps is hidden in composite subworkflows.} 
\label{fig:pip-table}
\end{figure}
\end{center}

\section{Data Organisation, Selection, and Routing}

The processing components in Kepler workflows are called actors. The Reflex package currently includes twelve actors for the development of data reduction workflows. Moreover, all other existing Kepler actors can be used. Four Reflex actors are usually involved  before the actual execution of data reduction recipes:

The {\bf DataOrganizer} organizes and classifies raw and reduced data: it takes as inputs all the files contained in a given directory, classifies them and organizes them into Lists of Science Observations. Each list contains science frames, all the associated calibrations needed to reduce them, and custom parameters for the reduction recipes. The user can check and modify the selection.

The {\bf DataSetChooser} allows the user to view and select the groups of files created by the DataOrganizer: it displays a list of LoSOs and provides
 the possibility of selecting, deselecting and analysing them. It produces one token on the output port per selected LoSO.

The {\bf FitsRouter} sorts files based on their category and routes them to the corresponding output port. The {\bf SofCombiner} groups related set of files into a common channel. For example, the averaged master bias together with the master flat, the order location table, the arc lamp exposure and the arc line catalog are routed to the wavelength calibration recipe. 

\section{CPL Recipe Executer}

\begin{center}
\begin{figure}[!ht]
\includegraphics[width=130mm,height=80mm]{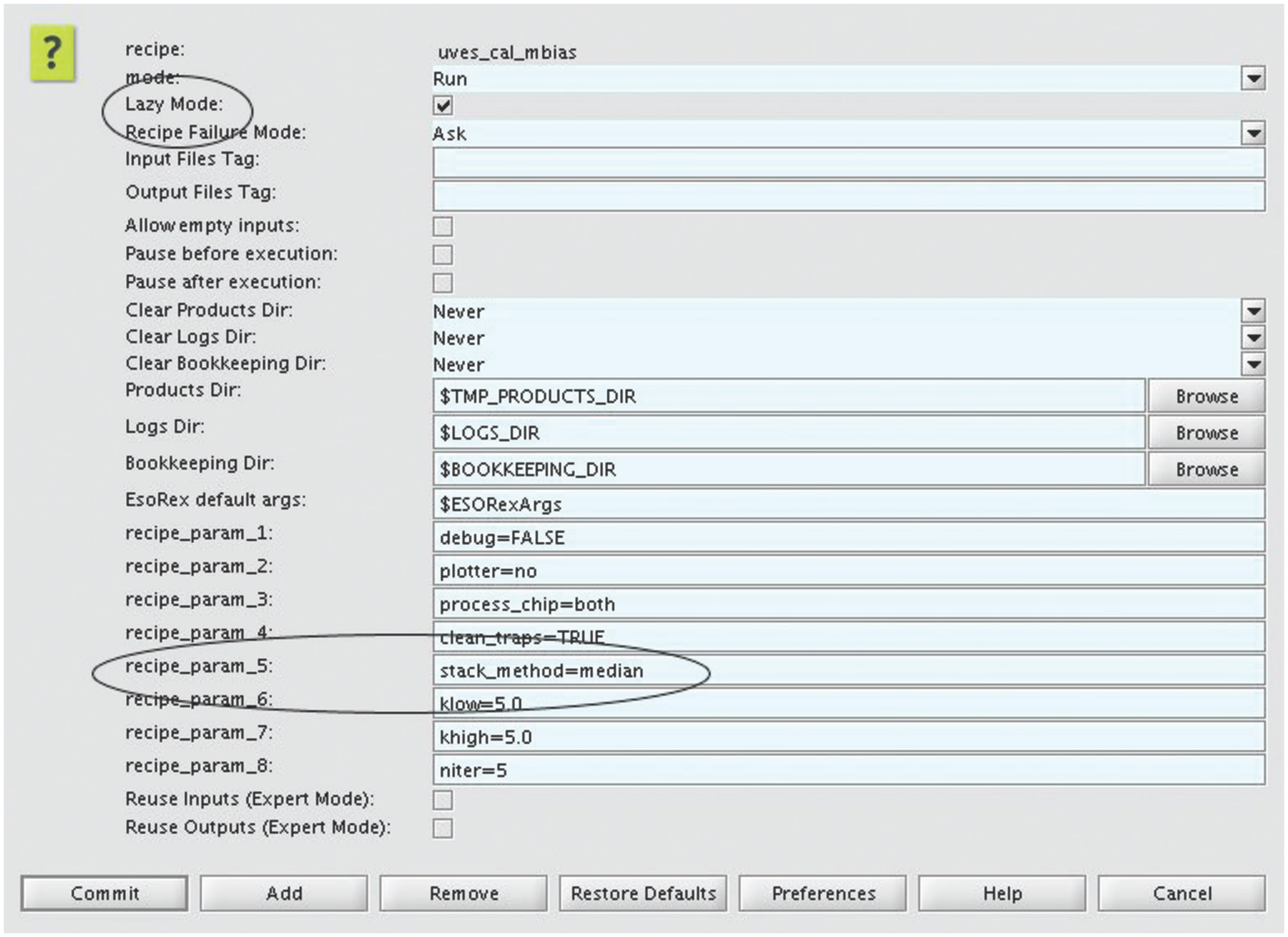}
\caption{The configuration panel for the RecipeExecuter actor. The list of parameters is created for a given pipeline recipe at the time the actor is instantiated in the workflow. All parameters of a recipe can be modified at this level. In addition the RecipeExecuter supports several execution modes, giving the user control over the input, output, and conditions of execution of the recipe.} 
\label{fig:pip-table}
\end{figure}
\end{center}

The RecipeExecuter executes a CPL recipe from the VLT pipelines (Fig. 2). In the workflow, each parameter of the recipe is initially set to the default value. The user can change any of the parameters, or provide calculated parameter values to a special input port. The RecipeExecuter supports a variety of execution modes, facilitating the processing of large data sets.

The Lazy mode is particularly useful: if true the RecipeExecuter will check whether the pipeline recipe has already been executed with the same input files and with the same recipe parameters. If this is the case then the recipe will not be executed, and instead the previously generated products (which are the same as those that would have been generated if the recipe were executed, except for a timestamp) will be broadcast to the output port. 

Kepler supports subworkflows, which make it possible to display the data processing tasks at different levels of detail. Examples of subworkflows are the iterative subworkflows: it is sometimes useful to be able to visualise the products of a recipe execution, tweak some recipe parameter and execute the recipe again, until the products are as expected. This can be easily achieved in Reflex by means of the RecipeLooper and the PythonActor, embedded in a Kepler subworkflow.

\section{Python Actor}

\begin{center}
\begin{figure}[!ht]
\includegraphics[width=130mm,height=80mm]{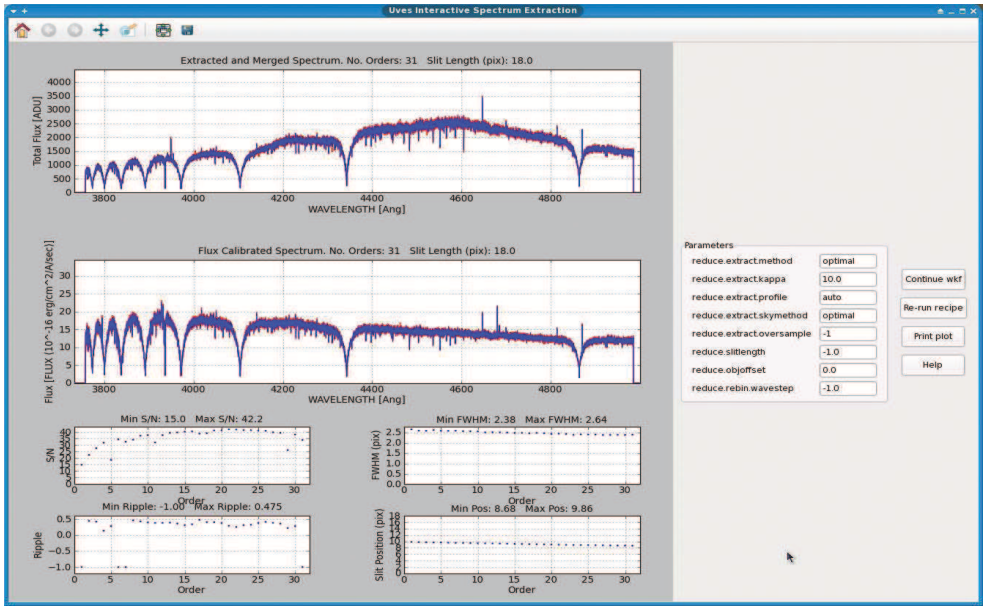}
\caption{The Python actor can be used to develop interactive windows that display intermediate results, and let the user fine-tune a set of recipe parameters.} 
\label{fig:pip-table}
\end{figure}
\end{center}

The PythonActor executes custom Python scripts. A module included in the Reflex release provides the functions necessary to insert scripts into the workflows. Upon selecting a script, input and output ports are automatically created, together with a parameter for each output port. Python actors are used to create plotting windows, graphical user interface, or to invoke tasks from external data reduction systems. The released UVES workflow comes with several examples of interactive Python actors  (Fig. 3).

\section{Where can I find Reflex?}

Reflex is distributed on the ESO Web site http://www.eso.org/reflex. The pipeline specific workflows, like the UVES workflow shown in this article are available from the ESO pipelines Web site: http://www.eso.org/pipelines, together with documentation, installation procedures, and test data.

\end{document}